\documentclass{emulateapj}
\usepackage{natbib}
\newcommand{\Msun}{{\rm M_{\odot}}}

\newcommand{\mpc}{\, {\rm Mpc}}
\newcommand{\kpc}{\, {\rm kpc}}
\newcommand{\pc}{\, {\rm pc}}

\slugcomment{Accepted for publication in ApJ Letters}

\shorttitle{Ultra Diffuse Galaxies in Coma Cluster}
\shortauthors{Jin Koda}

\begin{document}

\title{Approximately A Thousand Ultra Diffuse Galaxies in the Coma cluster}

\author{Jin Koda\altaffilmark{1},
Masafumi Yagi\altaffilmark{2,3},
Hitomi Yamanoi\altaffilmark{2},
Yutaka Komiyama\altaffilmark{2,4}}

\email{jin.koda@stonybrook.edu}

\altaffiltext{1}{Department of Physics and Astronomy, Stony Brook University, Stony Brook, NY 11794-3800}
\altaffiltext{2}{Optical and Infrared Astronomy Division, National Astronomical Observatory of Japan, 2-21-1 Osawa, Mitaka, Tokyo, 181-8588, Japan}
\altaffiltext{3}{Department of Advanced Sciences, Hosei University, 3-7-2, Kajinocho, Koganei, Tokyo, 184-8584 Japan}
\altaffiltext{4}{SOKENDAI (The Graduate University for Advanced Studies), Mitaka, Tokyo, 181-8588, Japan}

\begin{abstract}
We report the discovery of 854 ultra diffuse galaxies (UDGs) in the Coma cluster using deep $R$ band images,
with partial $B$, $i$, and H$\alpha$ band coverage,
obtained with the Subaru telescope.
Many of them (332) are Milky Way-sized with very large effective radii of $r_{\rm e}>1.5\kpc$.
This study was motivated by the recent discovery of 47 UDGs by
\cite{van-Dokkum:2015lr}; our discovery suggests $>1,000$ UDGs
after accounting for the smaller Subaru field ($4.1 \rm\, degree^2$; about 1/2 of Dragonfly).
The new Subaru UDGs show a distribution concentrated around the cluster center,
strongly suggesting that the great majority are (likely longtime) cluster members.
They are a passively evolving population, lying along the red sequence in the color-magnitude diagram
with no signature of H$\alpha$ emission.
Star formation was, therefore, quenched in the past.
They have exponential light profiles,
effective radii $r_{\rm e}\sim800\pc$-$5\kpc$, effective surface brightnesses $\mu_{\rm e}(R)=$25-28 $\rm mag\,arcsec^{-2}$,
and stellar masses $\sim 1\times 10^7\Msun$ - $5\times 10^8\Msun$.
There is also a population of nucleated UDGs.
Some MW-sized UDGs appear closer to the cluster center than previously reported;
their survival in the strong tidal field, despite their large sizes, possibly indicates a large dark matter fraction
protecting the diffuse stellar component. The indicated baryon fraction $\lesssim 1\%$ is less than
the cosmic average, and thus the gas must have been removed (from the possibly massive dark halo).
The UDG population is elevated in the Coma cluster compared to the field, indicating that the gas removal mechanism is
related primarily to the cluster environment.
\end{abstract}


\section{Introduction}

This study is motivated by the discovery of 47 ultra diffuse galaxies (UDGs) in the Coma cluster
by \citet{van-Dokkum:2015lr} using the Dragonfly Telescope Array \citep[][ hereafter Dragonfly]{Abraham:2014lr}.
This unexpected discovery revealed a new populatoin of low surface brightness (SB) galaxies.
Indeed, their central SBs are very low 24-26$\rm\, \,mag\,arcsec^{-2}$ in $g$-band and their median stellar mass
is only $\sim 6\times 10^7 \Msun$, despite their effective radii $r_{\rm e}=1.5$-$4.6 \kpc$ being as large as
those of $L_{\rm *}$ galaxies \citep[e.g., $\sim 3.6\kpc$ for the Milky Way (MW), calculated from ][]{Rix:2013lr}.
\citet{van-Dokkum:2015lr} speculated that the UDGs probably have very high dark matter fractions
as they have survived in the strong tidal field of the cluster.

Dragonfly is optimized to efficiently discover faint structures over a large field of view,
but has relatively poor spatial resolution with seeing and pixel scales
of $\sim 6\arcsec$ and $2.8\arcsec$, respectively.
The above properties of the Dragonfly UDGs were derived after their discovery from archival Canada France
Hawaii Telescope imaging.
Follow-up studies are needed to understand their nature and origin,
as well as their relationship to the cluster environment \citep[][ for review]{Boselli:2014fk}
and to other more studied galaxy populations, such as dwarfs and low SBs, in clusters
\citep[e.g., ][]{Binggeli:1991fk, Bothun:1991kx, Ulmer:1996vn, Ulmer:2011yq, Adami:2006yq,Adami:2009uq, Lieder:2012qy,Ferrarese:2012lr}.

Optical telescopes of larger aperture are advantageous for a resolved study of this population.
\cite{Yamanoi:2012fk} used the Subaru Prime Focus Camera \cite[Suprime-Cam; ][]{Miyazaki:2002fk}
on the Subaru telescope and derived a galaxy luminosity function down to $M_{\rm R}\sim-10$ in Coma.
Their three $34\arcmin \times 27\arcmin$ fields include nine Dragonfly UDGs.
All of the nine were easily found in their catalog, being resolved spatially in the images.
Therefore, Subaru imaging can identify this new galaxy population efficiently and
permits an investigation of their internal properties.
Several archival Subaru images are available for the Coma cluster \citep{Yagi:2007uq, Yoshida:2008kx, Yagi:2010ve, Okabe:2010yq, Okabe:2014fj}.
In this Letter, we use the archival Subaru data and report the discovery of 854 UDGs, implying $\sim 1000$ UDGs after scaling
for the Dragonfly field-of-view.

We adopted $m-M$ = 35.05 \citep{Kavelaars:2000rt} as the distance modulus of the Coma cluster,
which corresponds to an angular diameter distance of $97.5\mpc$ ($1\arcsec = 0.47\kpc$)
\footnote{We adopted the Cosmological parameters of ($h_0$,$\Omega_M$, $\Omega_{\lambda}$)=(0.71, 0.27,0.73) from \cite{Larson:2011lr}.}.
The full catalog of the Subaru UDGs will be published in Yagi et al. (in preparation).
We use the AB-magnitude system in this work.

\section{Data}

\begin{figure*}
\epsscale{1.2}
\plotone{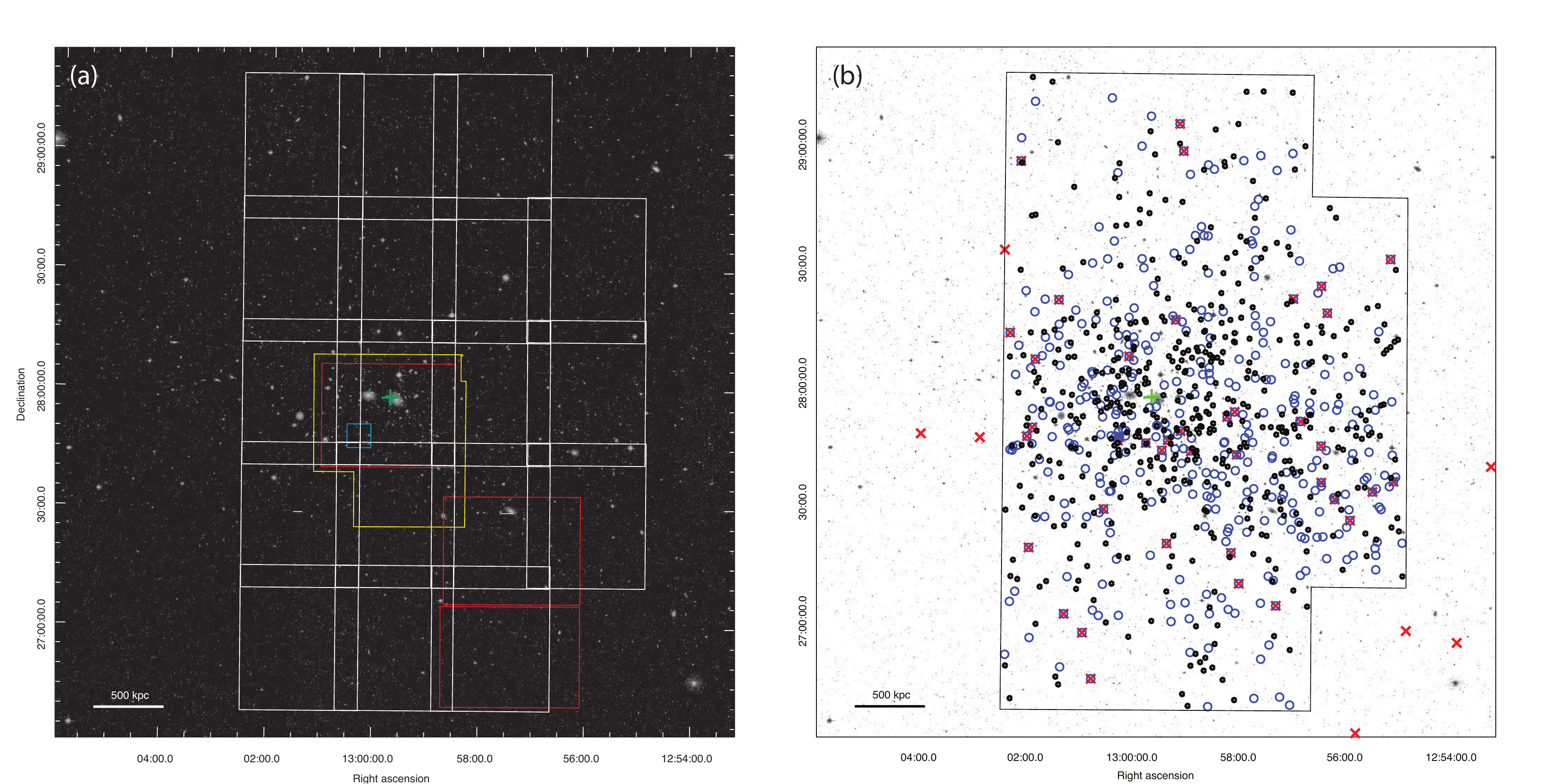}
\caption{The $2.86\deg \times 2.90\deg$ ($\sim4.87 \times 4.94\mpc^2$) area centered on the Coma cluster,
the same area as in Figure 1 of \citet[][]{van-Dokkum:2015lr}.
(a) Image from the Digitized Sky Survey.
The white borders show the 18 fields covered in the Subaru $R$ band \citep{Okabe:2014fj},
which have the total area of $4.1 \rm\, degree^2$, about 1/2 of the Dragonfly coverage.
Red indicates the area analyzed by \cite{Yamanoi:2012fk}.
Yellow outlines the area analyzed by \citet{Yagi:2010ve} using the Subaru $B$, $R$, H$\alpha$, $i$ bands.
Cyan indicates the area in Figure \ref{fig:colorimage}.
The center of the cluster ($\alpha_{\rm J2000}$,$\delta_{\rm J2000}$)=(12:59:42.8,+27:58:14) is marked with a green cross \citep{White:1993lr}.
(b) The same area as in (a), showing the distribution of the 854 Subaru UDGs (circles).
The MW-sized UDGs, with large effective radii ($>1.5\kpc$), are shown in blue.
The Subaru field coverage in $R$ is enclosed with the solid line.
The 47 Dragonfly UDGs are indicated with red crosses. 
}\label{fig:coverage}
\end{figure*}

The raw $R$ band images from the Suprime-Cam were
obtained from the Subaru data archive \citep{Baba:2002qy}.
Suprime-Cam has a mosaic of ten $2048 \times 4096$ CCDs and
covers a wide field of $34\arcmin \times 27\arcmin$ with a pixel scale of $0.202\arcsec$.
An eighteen-pointing mosaic with Suprime-Cam was taken by \citet{Okabe:2014fj}
and covered about $4.1 \deg^2$ (Figure \ref{fig:coverage}).
The seeing was $0.6$-$0.8\arcsec$, typically $0.7\arcsec$.
Integration times for the 18 fields were not the same, resulting in variations in background noise, 
i.e., $28.3$-$28.7\,\rm mag\,arcsec^{-2}$ (1$\sigma$) in a $2\arcsec$ aperture
(equivalent to $30.0$-$30.4\,\rm mag\,arcsec^{-2}$ in a $10\arcsec$ aperture, $\sim1$ mag deeper than \citealp{van-Dokkum:2015lr}).
The very central field has a higher variation of $27.8\,\rm mag\,arcsec^{-2}$ since the field is contaminated
by the outer envelope of bright galaxies.

The data were reduced in a standard way \citep{Yagi:2002lr, Yagi:2010ve}.
We used self-sky flat images, subtracted sky background locally in each small grid ($256\times256\,\rm pixel^2$;
$51.7\arcsec \times 51.7\arcsec$), used the WCSTools \citep{Mink:2002fk} for astrometry,
and applied a photometric calibration \citep{Yagi:2013fk} using  the Sloan Digital Sky Survey (SDSS)-III DR9 catalog \citep{Ahn:2012kx}.
The grid size for the background subtraction was  larger than the expected size of UDGs ($<30\arcsec \sim 15\kpc$).
The Galactic extinction in $R$ band varies from 0.016 to 0.031 mag across the 18-field mosaic \citep{Schlafly:2011qy}.
We adopted the Galactic extinction value at the center of each field and
neglected variation within each field.
The final photometric error is $\lesssim 0.1$ mag.
More details of the data reduction procedure will be presented in Yagi et al. (in preparation).
In addition, we used the Suprime-Cam $B$, $i$, and H$\alpha$ reduced images (see Figure \ref{fig:coverage})
by \citet{Yagi:2010ve} and \citet{Yamanoi:2012fk}.

We also analyzed a control field for comparison.
The $R$ band data of one Suprime-Cam pointing, 1/18 of the Coma field,
were taken from the Subaru Deep Field (SDF) project \citep{Kashikawa:2004uq}.
We used only a part of the raw SDF exposures taken in June 2008 to make the background noise 
comparable to that in the Coma fields. The 1$\sigma$ background noise in a $2\arcsec$ aperture
is $28.6\,\rm mag\,arcsec^{-2}$.
For consistency we started from the raw data and matched data reduction parameters.

\begin{figure*}
\epsscale{1.1}
\plotone{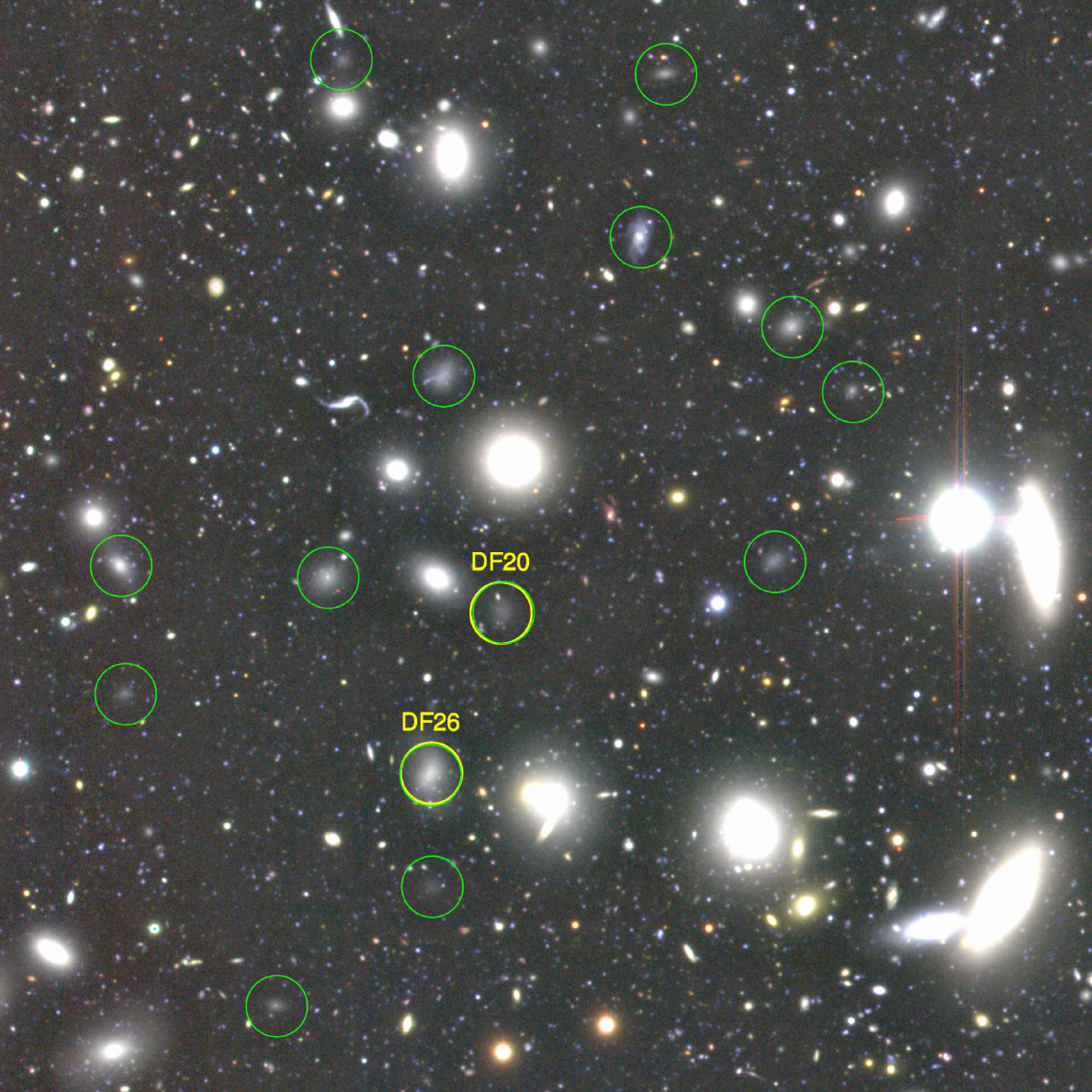}
\caption{
Subaru $BRi$ color image of the $\sim6\arcmin \times 6\arcmin$ region ($\sim 170 \times170 \kpc^2$ region at $d=97.7\mpc$), shown in cyan in Figure \ref{fig:coverage}a.
The Dragonfly and Subaru UDGs are marked respectively with yellow and green circles with a diameter of $20\arcsec$ ($\sim 9.5\kpc$).
}\label{fig:colorimage}
\end{figure*}

\section{Identification}

Our goal is to find UDGs in the Subaru images.
Forty of the 47 UDGs discovered by Dragonfly are within the Subaru $R$ band coverage
based on their coordinates \citep{van-Dokkum:2015lr}.
All were detected significantly (with the faintest one, DF27, off by $12.5\arcsec$ from the published coordinate)
and their structures were resolved in the Subaru images.
The detection threshold was approximately 27.3 $\rm \, mag\, arcsec^{-2}$ in the $R$ band.
We describe our selection procedure for the final catalog of 854 UDGs in the Coma cluster.
We found no counterparts in the control field.

We ran SExtractor \citep[version 2.19.5; ][]{Bertin:1996lr} on individual mosaic frames
with a fixed detection threshold of 27.3 $\rm \, mag\, arcsec^{-2}$ in $R$.
We removed a first set of spurious detections using SExtractor's "$\rm FLAGS<4$" and "$\rm PETRO\_RADIUS>0$".
This initial catalog had 2,627,495 objects, including duplicates in the overlap regions of adjacent mosaic frames ($\sim 30$ \%).
We used the Dragonfly UDGs as the fiducial set in adjusting parameters for selection of UDG candidates,
but could not use exactly the same selection criteria as \citet{van-Dokkum:2015lr} due to the difference in image quality.
We applied constraints on $R$ magnitude and size, "$\rm 18<MAG\_AUTO<26$" and "$\rm FWHM(Gaussian)>4\arcsec$"
(i.e., all Dragonfly UDGs satisfy this condition), which left 7,362 objects.

The reported effective radius of the Dragonfly UDGs is $r_{\rm e}\gtrsim3.2\arcsec$ \citep[using $r_{\rm e}$ from GALFIT; ][]{Peng:2002fk}.
However, in the resolved Subaru images we found that an alternative constraint, SExtractor's $r_{\rm e}\gtrsim1.5\arcsec$,
captures all the Dragonfly UDGs.
We therefore used $r_{\rm e}>1.5\arcsec$ and a mean SB of $\left< \mu(r_{\rm e}) \right> >24$
to choose UDG candidates. [Note that we found that $r_{\rm e}$ from SExtractor and GALFIT were occasionally very inconsistent;
we use SExtractor's $r_{\rm e}$ for identification and GALFIT's $r_{\rm e}$ for studies of structural properties.]
We excluded objects with high central concentrations (mostly high-z galaxies) by removing
those whose mean SB within $r_{\rm e}$ deviates largely from the SB
at $r_{\rm e}$. This constraint, $\mu_{\rm e}-\left< \mu(r_{\rm e}) \right><0.8$,
left 1,779  candidates.

The final step was removal of spurious objects by visual inspection.
Most spurious detections were due to the crowding in the cluster, such as faint tidal tails and galaxy blending,
as well as distant edge-on disk galaxies, artifacts at image edges, and optical ghosts.
To minimize human error, the four authors separately went through all postage stamp images.
After this step and removal of duplications based on their coordinates,
854 UDG candidates were left on which at least three of us agreed.
The full catalog will be published by Yagi et al. (in preparation).

\section{Ultra-Diffuse Galaxy Candidates}

The 854 UDGs candidates from Subaru are visually comparable to the Dragonfly UDGs.
Figure \ref{fig:colorimage} shows a sample $6\arcmin \times 6\arcmin$ field,
showing the Subaru (green circles) and Dragonfly UDGs (yellow).
Their low SBs are evident compared to the surrounding galaxies,
including major galaxies in the cluster and distant background ones.
Their large sizes are also clear when compared to the $20\arcsec$ diameter of the circles
($\sim 9.5\kpc$ at $d=97.5\mpc$).
The greater number of detections, compared to Dragonfly, may be due to
the superior seeing (less blending) and higher signal-to-noise ratio.

The majority of the 854 candidates are most likely UDGs in the Coma cluster.
One of them has been spectroscopically confirmed as a cluster member \citep{van-Dokkum:2015fk}.
The control SDF field has virtually no counterparts --
only thirteen were left after the SExtractor-based
selection, twelve of which were obvious image artifacts or tails of bright galaxies.
The last one appeared to be a blend of multiple objects.
Hence, contamination by non-cluster members is rare and negligible.
Note, however, that some negligible number of contaminations might still exist.
For example, the third object from the top in Figure \ref{fig:colorimage}
may be a background spiral galaxy.
Despite this significantly increased sample, the UDGs are still a minor population
in the Coma cluster \citep{Yamanoi:2012fk}.

In the literature, we found that many of the Subaru UDGs
had been cataloged albeit as more compact objects;
\citet{Adami:2006kx} found 248 of 309 that lay within their coverage, and  \citet{Yamanoi:2012fk} 232 of 240.
Among them, only 17 were classified as low SB galaxies \citep{Adami:2006yq}.
Their large extents and low SBs were revealed for the first time in this study.
We note that eleven out of the 12 Dragonfly UDGs  within their field were
also cataloged in \citet[][]{Adami:2006kx}, but none were classified as low SB \citep{Adami:2006yq}.

\section{Structural Parameters}\label{sec:param}

\begin{figure*}
\epsscale{1.1}
\plotone{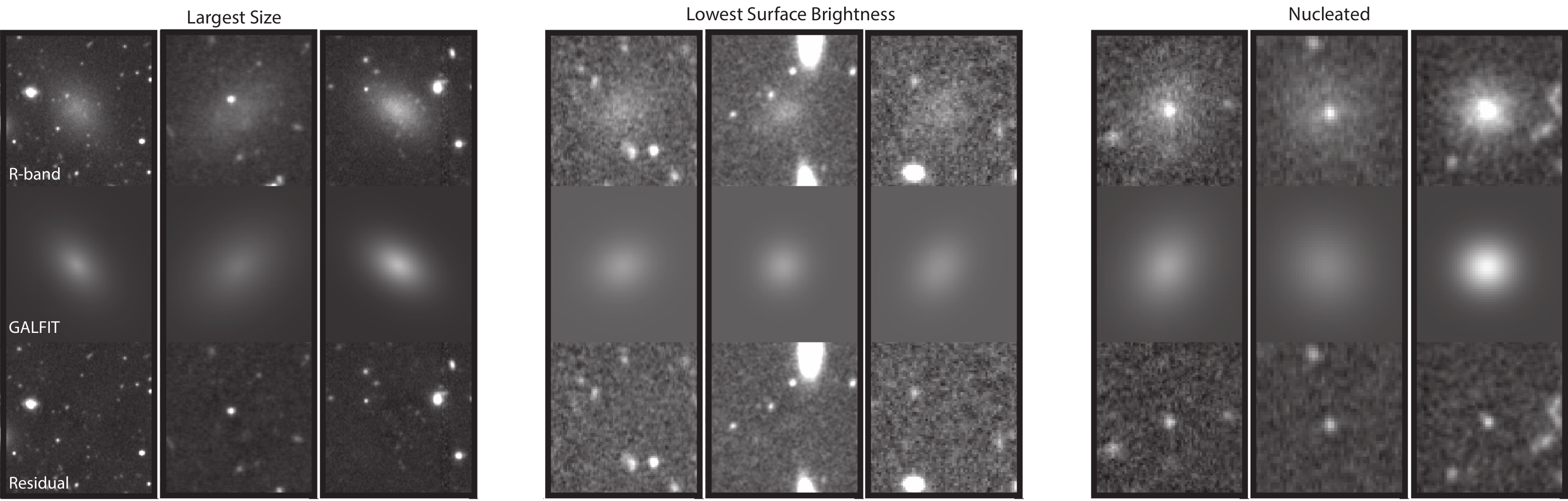}
\caption{Examples of GALFIT results drawn from the groups of largest-size UDGs, lowest surface-brightness UDGs, and nucleated UDGs.
}\label{fig:galfit}
\end{figure*}

The GALFIT package was used to measure the structural parameters of the Subaru UDGs.
The fits were made with a single Sersic profile \citep{Sersic:1968uq} with sky background fixed.
We used SExtractor's segmentation images to mask surrounding objects.
Seventy-nine of the 854 required additional manual masks to exclude a bright compact object(s)
within their boundaries -- interestingly, sixty-seven of these appear to have compact nuclei at their centers.
We judged the fits acceptable based on the goodness-of-fit ($\chi^2_{\nu}<1$; 75 objects were thus excluded)
and consistency between GALFIT and SExtractor measurements ($r_{\rm e}$ from GALFIT and
SExtractor consistent within a factor of 3; 11 removed).
In this section, we use the sample of 768 objects with good GALFIT results, out of which 332
have GALFIT's $r_{\rm e}>1.5\kpc$ \citep[i.e., MW-sized UDGs as defined in ][]{van-Dokkum:2015lr}.
We refer to the former full set of galaxies as UDGs, and the latter as MW-sized UDGs.
Figure \ref{fig:galfit} shows some examples of GALFIT results for the UDGs of lowest SB,
of largest-size, and with a compact nucleus.

\begin{figure*}
\epsscale{1.1}
\plotone{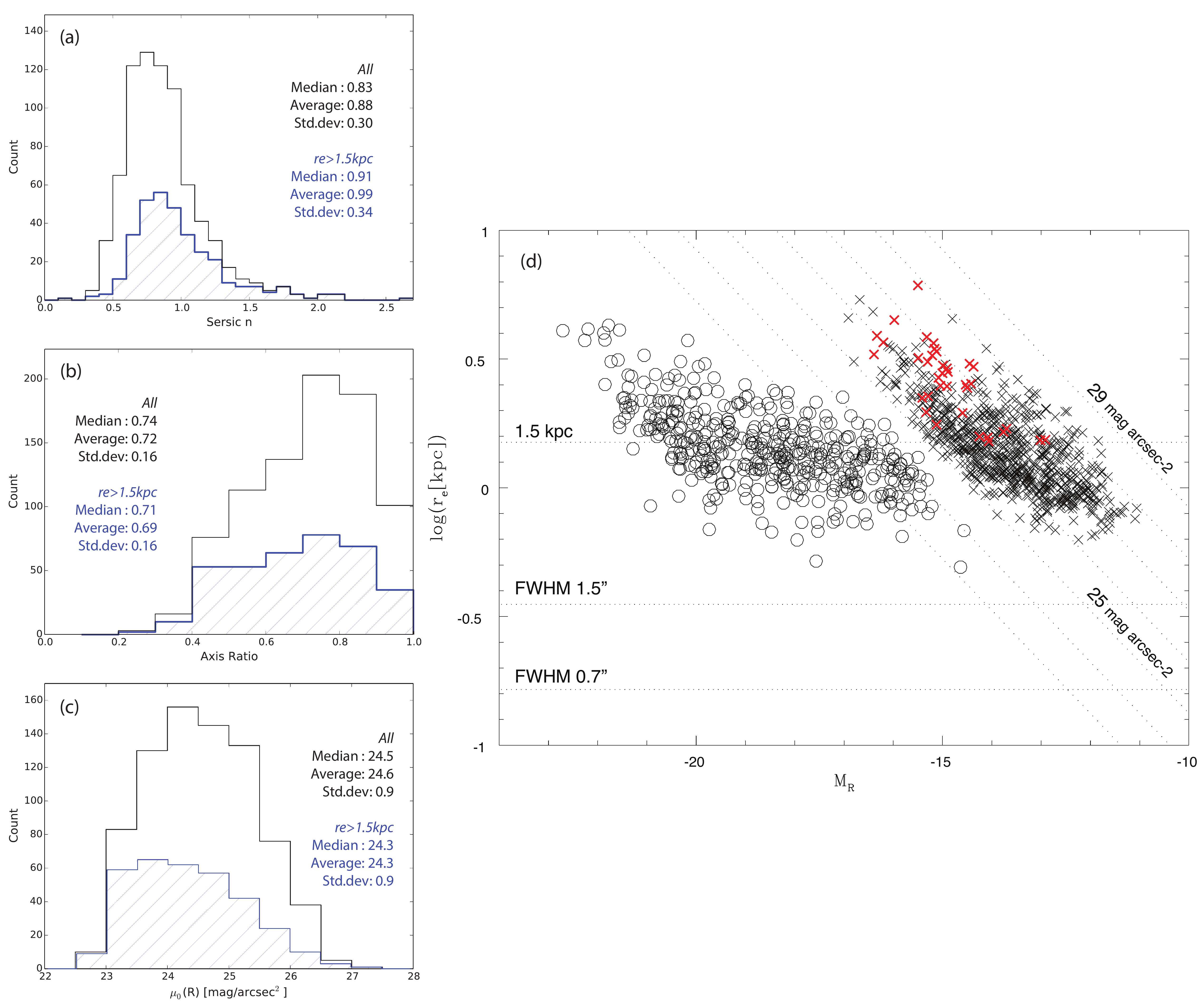}
\caption{
Structural properties of UDGs.
(a) Histograms of Sersic index $n$, (b) axis ratio $b/a$, and (c) central SB $\mu_0{\rm (R)}$
with their medians, averages, and standard deviations.
Black lines are for all 854 UDGs, while blue are for 332 MW-sized UDGs alone.
(d) Effective radius vs. $R$ magnitude.
The parameters of the UDGs (crosses; red for the Dragonfly UDGs) are derived with GALFIT.
Normal galaxies (circles) --spestroscopically-confirmed Coma members \citep{Mobasher:2001fj} --
are also plotted for comparison \cite[from ][ with the conversion R(AB)-R(Vega)=0.21]{Komiyama:2002uq}.
Dotted, diagonal lines show constant SBs ($\mu_{\rm e}$s) from 23 to 29 mag arcsec$^{-2}$ with
a 1 mag arcsec$^{-2}$ interval for the case of an exponential profile (note $\mu_0=\mu_e-1.82$ for $n=1$).
The gap between the normal galaxies and
UDGs is due to selection effects.
Horizontal lines show $r_e$ of PSF
with a FWHM of 1.5 arcsec (Komiyama et al. 2002)
and a FWHM of 0.7 arcsec (this study).
}
\label{fig:stat}
\end{figure*}

Figure \ref{fig:stat}(a), (b), and (c) show histograms of Sersic index ($n$), axis ratio ($b/a$), and central SB ($\mu_0$(R)).
For both UDGs and MW-sized UDGs, their average Sersic indices $\left< n\right>=0.9$-$1.0$ indicate
an exponential profile. The distributions of axis ratio, as well as its average $\left<b/a\right>=0.7$-$0.8$,
are skewed toward a large value; therefore, this UDG sample does not consist of randomly-oriented thin-disk galaxies
in a statistical sense (which would skew their distribution toward a low $b/a$).
The $\mu_0$(R) ranges around 23-26$\,\rm mag\,arcsec^{-2}$.
These results are consistent with \citet{van-Dokkum:2015lr} when the difference in the adopted bands, SDSS~$g$ and
Subaru~$R$, is taken into account (roughly $g-R\sim$0.8 mag for the red-sequence in Coma).

The Subaru UDGs are likely an extension of normal and dwarf galaxy populations and are not,
on their own, a distinct population.
Figure  \ref{fig:stat}(d) shows the properties of the UDGs (crosses) with respect to normal galaxies
in Coma \citep[circles; from ][]{Komiyama:2002uq}.
The apparent $R$ magnitude of the UDGs is 18 to 24 mag, indicating an absolute magnitude of
about -12 to -16 mag at the Coma distance.
The smallest and faintest UDGs (e.g., $r_{\rm e}<1\kpc$ and $M_{\rm R}<-12$ in Figure \ref{fig:stat}c)
overlap with the largest and brightest dwarf galaxies 
and share some properties in common with them
\citep[e.g., the exponential profile, nucleated population; see ][]{Tolstoy:2009lr, McConnachie:2012fk, Boselli:2014fk}.
Dotted lines represent constant SBs ($\mu_{\rm e}$s)
from 23 to 29 $\rm mag\, arcsec^{-2}$ with a 1 $\rm mag\, arcsec^{-2}$ interval, assuming an exponential
profile. The average SB of the Subaru UDGs is distributed from about 25 $\rm mag\, arcsec^{-2}$
(i.e., a cut-off due to the selection) to 28 $\rm mag\, arcsec^{-2}$ (due to the detection limit; 
this lower boundary is lower than the pix-to-pix detection limit, because the UDGs are extended).

The absolute magnitudes correspond to stellar masses of $1\times10^7\Msun$-$5\times10^8\Msun$
if we adopt a mass-to-light ratio of $M/L_{\rm R}\sim 3$.
Note the $M/L_{\rm R}$ varies by a factor of $\sim2$ for ages of 4-12 Gyr and
metallicities between 0.2-1.0 solar based on calculations using
Starburst99 \citep{Leitherer:1999qh}, a single starburst, and a Kroupa initial mass function.

\section{A Passively-Evolving Population}\label{sec:passive}

The UDGs are distributed widely over the entire area of the cluster with a concentration
toward its center (Figure \ref{fig:coverage}b).
This spatial correlation also supports the assumption that
the great majority are cluster members.
Figure \ref{fig:coverage} nearly covers the virial radius of the cluster
\cite[$\sim2.8$ Mpc; $\sim 1.7\deg$; ][]{Kubo:2007lr},
and reveals their relatively symmetric distribution around the center with a potential
elongation toward the south west (roughly toward NGC 4839).
This symmetric, wide-spread distribution may indicate their long history within the cluster.

\begin{figure*}
\epsscale{0.8}
\plotone{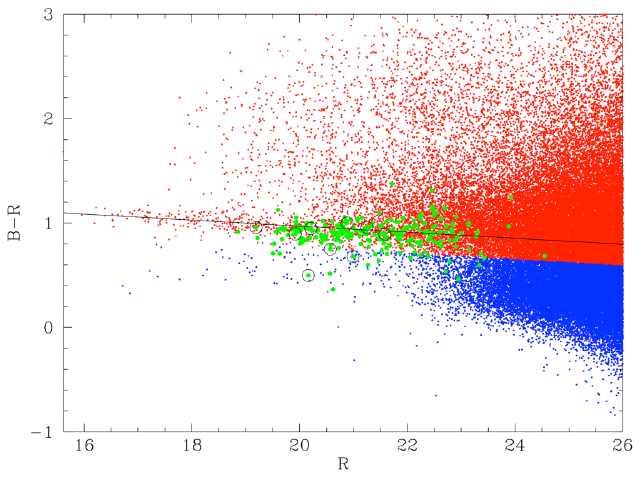}
\caption{Color-magnitude diagram using $B$ and $R$ band photometry.
The green points are 232 UDGs observed both in $B$ and $R$ with Subaru
(the Dragonfly UDGs are circled),
and the red and blue are red and blue galaxies taken from the Coma1 field of \citet{Yamanoi:2012fk}
which includes cluster members as well as background galaxies.
Due to saturation, most giant galaxies are not included, but the red-sequence is evident.
The UDGs clearly follow the red-sequence population of the Coma cluster.
}\label{fig:cmd}
\end{figure*}

The UDGs closely follow the red-sequence of a passively-evolving galaxy population on
the color-magnitude diagram.
232 UDGs are in the catalog of \citet{Yamanoi:2012fk}
with both $B$ and $R$ photometry.
Figure \ref{fig:cmd} shows their distribution (green).
The comparison data (red and blue) show other cluster member galaxies, as well as
background galaxies, and are also from \citet{Yamanoi:2012fk}.
The red-sequence is evident, and the solid line is a fit by \cite{Yamanoi:2012fk}.
Clearly, the UDGs lie along this red-sequence, and
their $B$-$R$ colors are around 0.8-1.0 mag.
This is similar to the trends found among dwarfs and low SB galaxies in clusters
\citep{Adami:2009uq,Ulmer:2011yq,Lieder:2012qy}.

No significant H$\alpha$ excess was found in UDGs.
217 UDGs are within the Subaru H$\alpha$ coverage (yellow in Figure \ref{fig:coverage}a),
which was designed to detect faint H$\alpha$ emission around Coma member galaxies \citep{Yagi:2010ve}.
Therefore, the UDGs are not forming stars at the current epoch, as expected for passively-evolving galaxies.

\section{Discussions}\label{sec:discussion}

We report the discovery of $\sim1000$ UDGs in the Coma cluster, about 40\% of which are MW-sized.
The new UDG sample is by no means complete, but already contains 10-20 times more
than previously known \citep{van-Dokkum:2015lr}.
None of the UDGs show a signature of tidal distortions; this is our selection criteria,
but indicates that this sample of UDGs are not likely recently-disrupted tidal debris.

The UDG population, compared to brighter galaxies, is elevated in the Coma cluster,
although a small number of large, low-SB galaxies are known
in the field \citep{Dalcanton:1997qy, Impey:2001ys, Burkholder:2001fr}.
f the UDG-to-brighter galaxy number ratio in Coma were common in the field, 
the expected UDG population would be implausibly large, $\gtrsim 10^5$ within 100 Mpc of the MW.
To obtain this rough estimate, we used, as a reference, galaxies in the SDSS \citep{Ahn:2012kx}
within $16<r<17$ mag
($-19<M_{r}<-18$ in absolute magnitude) at the cluster's redshift of 0.013-0.033 \citep{Mobasher:2001fj}.
The number of reference galaxies in the field was estimated from the luminosity function of \citet{Blanton:2001pz}.
The estimated number in the field, $\gtrsim 10^5$, is crude, but seems too large compared
to the small number discovered so far. The cluster environment must play a role in their formation and evolution.

\citet{van-Dokkum:2015lr} speculated that the MW-sized UDGs might be a dark-matter (DM) dominated
population in order for them to survive in the strong tidal field around the cluster core.
In fact, the Dragonfly UDGs spatially avoided the central $r\sim$300 pc region
as if the ones there had been tidally disrupted; this apparent disruption was
used to constrain the DM fraction \citep[as large as $\gtrsim 98\%$; ][]{van-Dokkum:2015lr}.
Surprisingly, we found UDGs even closer to the core (Figure \ref{fig:coverage}b),
and the closest one is MW-sized only about $3\arcmin$ ($\sim 85\kpc$)
away on the sky. Eleven UDGs were found within a radius of $5\arcmin$ ($\sim 141\kpc$).
These detections were, of course, not complete due to the high background emission there,
and their apparent proximity may result from a chance coincidence along a line of sight.
If any of them are within $\sim 100$-$150\kpc$ from the core, an even larger DM mass than
the estimate by  \citet{van-Dokkum:2015lr} is necessary,
and the baryon fraction within a tidal radius should be $\lesssim 1\%$.
This is below the cosmic average, and therefore, the baryons must have been removed from
the possibly very deep DM potential.

The possible removal of the gas, and quenching of star formation (SF), are consistent with their red color
and clustering in Coma (indicating their longevity within the cluster).
The red-sequence can be produced by a metallicity-sequence if galaxies have been evolving passively
since SF was quenched \citep{Kodama:1997rt}.
Physical processes often suggested for the quench include \citep[see ][ for review]{Boselli:2014fk}:
(a) blow out of gas due to galactic winds from supernovae or AGN activities  \citep{Dekel:1986lr, Arimoto:1987qy},
(b) ram-pressure stripping \citep{Gunn:1972lr},
(c) tidal-interaction and harassment \citep{Moore:1996lr}, and
(d) starvation due to the cessation of gas infall  \citep{Larson:1980qy}.
The elevated population in the cluster indicates that environmentally-driven mechanisms,
such as  (a), (b), and (c) are the most likely solutions [(a) may occur if SF
is induced by (b) or (c).].

\acknowledgments

We thank Alessandro Boselli, Samuel Boissier, Jim Barrett, and an anonymous referee for helpful comments.
This research utilized facilities and resources provided by the National Astronomical Observatory
of Japan (NAOJ), including the Subaru Telescope, the SMOKA data archive system, and
computers at the Astronomy Data Center.
JK acknowledges support from NASA grant NNX14AF74G and
NSF grant AST-1211680.





\end{document}